# Review of Deep Learning Applications to Structural Proteomics Enabled by Cryogenic Electron Microscopy and Tomography

Brady K. Zhou[1], Jason J. Hu[2,3], Jane K.J. Lee[4], Z. Hong Zhou[5,6,*], Demetri Terzopoulos[1]

1  Computer Science Department, University of California, Los Angeles, CA 90095, USA

2  Innovative Genomics Institute, University of California, Berkeley, CA 94720, USA

3  Department of Molecular and Cell Biology, University of California, Berkeley, CA 94720, USA

4  Department of Structural Biology, Stanford University, Stanford, CA 94305, USA

5  California NanoSystems Institute, University of California, Los Angeles, CA 90095, USA

6  Department of Microbiology, Immunology and Molecular Genetics, University of California, Los Angeles, CA 90095, USA

* Corresponding author

email: Hong.Zhou@UCLA.edu; Phone: 310-694-7527

## Abstract

**The past decade has witnessed a transformative "cryoEM revolution" characterized by exponential growth in high-resolution structural data, driven by advances in cryogenic electron microscopy (cryoEM) and cryogenic electron tomography (cryoET). The integration of deep learning technologies into structural proteomics workflows has emerged as a pivotal force in addressing longstanding challenges, including low signal-to-noise ratios, preferred orientation artifacts, and missing-wedge problems that have historically limited efficiency and scalability. This review article examines the application of Artificial Intelligence (AI) across the entire cryoEM pipeline, from automated particle picking using convolutional neural networks (Topaz, crYOLO, CryoSegNet) to computational solutions for preferred orientation bias (spIsoNet, cryoPROS) and advanced denoising algorithms (Topaz-Denoise). In cryoET, tools such as IsoNet employ U-Net architectures for simultaneous missing-wedge correction and noise reduction, while TomoNet streamlines subtomogram averaging through AI-driven particle detection. The workflow culminates with automated atomic model building using sophisticated tools like ModelAngelo, DeepTracer, and CryoREAD that translate density maps into interpretable biological structures. These AI-enhanced approaches have demonstrated remarkable achievements, including near-atomic resolution reconstructions with minimal manual intervention, resolution of previously intractable datasets suffering from severe orientation bias, and successful application to diverse biological systems from HIV virus-like particles to in situ ribosomal complexes. As deep learning continues to evolve, particularly with the emergence of large language models and vision transformers, the future promises even more sophisticated automation and accessibility in structural biology, potentially revolutionizing our understanding of macromolecular architecture and function.**

**Key Words: CryoEM, CryoET, Artificial intelligence, Deep learning, CNN, U-net, 3D Reconstruction**

## 1. Introduction

The past decade has witnessed a transformative "cryoEM revolution," driven by advancements in cryogenic Electron Microscopy (cryoEM) and cryogenic Electron Tomography (cryoET) [1-4]. Characterized by an exponential increase in high-resolution structural data, this revolution has not only been propelled by innovations in hardware, such as direct electron detectors, but also computational methods that enhance data processing and analysis, which has enabled an unprecedented level of detail in the determination of atomic-resolution structures of macromolecular complexes. However, challenges such as low Signal-to-Noise ratios (SNRs) and preferred orientation artifacts have historically limited the efficiency and scalability of cryoEM and cryoET workflows (Figure 1). The integration of *deep learning technologies* has emerged as a pivotal solution that automates critical tasks, improves resolution, and addresses longstanding bottlenecks in structural proteomics.

The deep learning approach to Artificial Intelligence (AI) uses deep neural networks to extract meaningful patterns from complex, high-dimensional datasets [5]. Such datasets are ideal for cryoEM and cryoET applications.

IsoNet [6] and TomoNet [7] are examples of tools that employ advanced neural network architectures such as U-nets and Convolutional Neural Networks (CNNs) to enhance particle picking, denoising, and 3D reconstruction. IsoNet corrects the "missing-wedge" problem in cryoET, producing isotropic tomograms, while TomoNet's AI-driven "AutoPicking" streamlines particle detection on flexible lattices. These advancements reduce cumbersome manual intervention, improve data quality, and achieve near-atomic resolutions. By addressing challenges such as noisy micrographs and orientation bias, deep learning has significantly expanded the scope and efficiency of structural biology research.

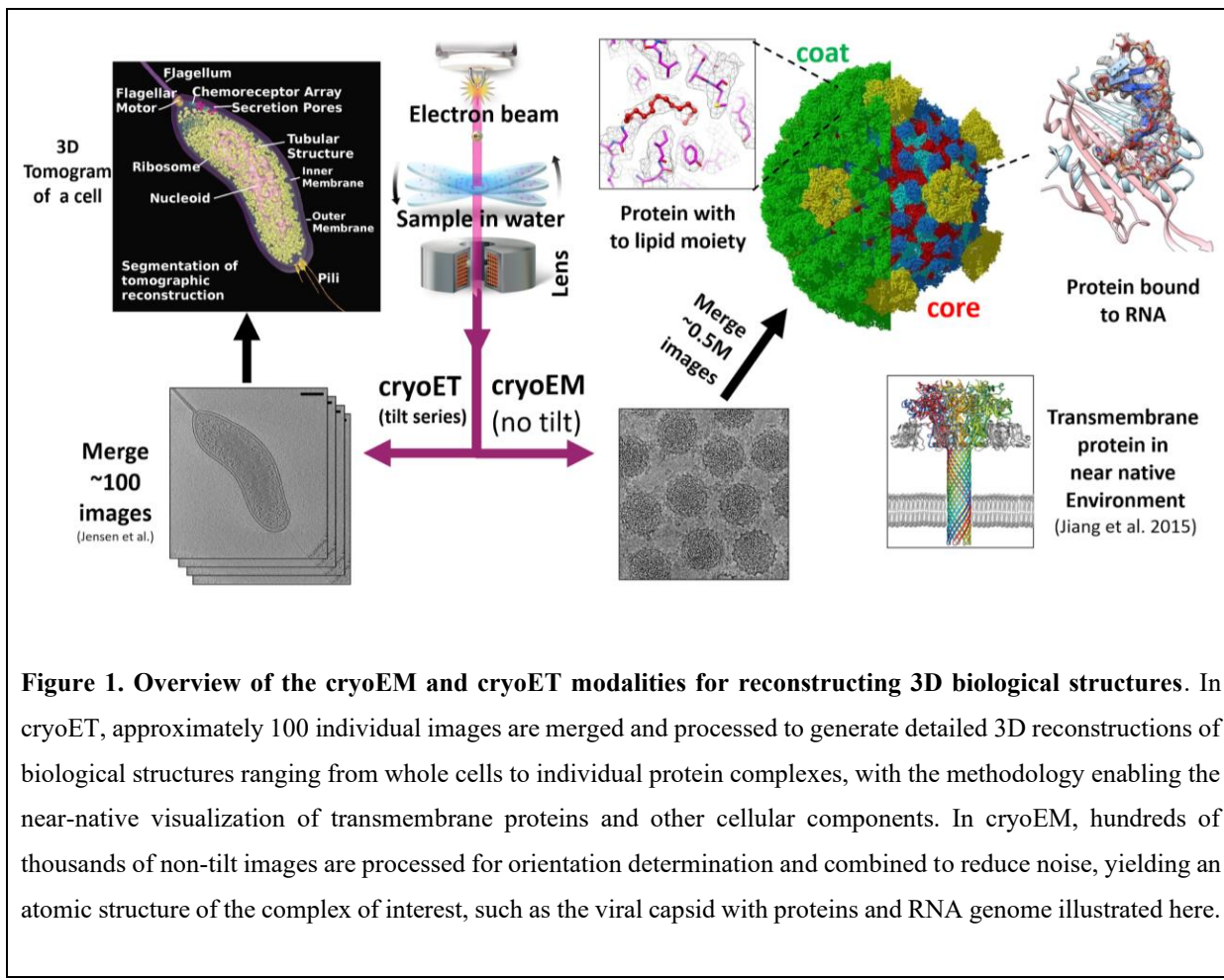

**Figure 1. Overview of the cryoEM and cryoET modalities for reconstructing 3D biological structures**. In cryoET, approximately 100 individual images are merged and processed to generate detailed 3D reconstructions of biological structures ranging from whole cells to individual protein complexes, with the methodology enabling the near-native visualization of transmembrane proteins and other cellular components. In cryoEM, hundreds of thousands of non-tilt images are processed for orientation determination and combined to reduce noise, yielding an atomic structure of the complex of interest, such as the viral capsid with proteins and RNA genome illustrated here.

The potential of AI in cryoEM and cryoET extends beyond current applications, with the explosion of AI models such as Large Language Models (LLMs) poised to further transform the field. These models could be fine-tuned on cryoEM datasets to automate complex tasks such as metadata interpretation or even synthetic data generation for model training. The success of hybrid approaches, such as CryoSegNet's integration of U-nets with "foundation models," particularly the Segment Anything Model (SAM), highlights the power of combining specialized and pre-trained architectures [8]. As deep learning continues to evolve, its integration into cryoEM workflows promises to make high-resolution structural analysis more accessible, reproducible, and efficient, paving the way for breakthroughs in understanding molecular structures and their biological functions.

3 | Page

This review provides a comprehensive examination of artificial intelligence applications across the entire cryoEM and cryoET pipeline (Figure 1), from initial data acquisition to final atomic model construction. The primary objective is to systematically review how *deep learning technologies* have addressed longstanding challenges in structural biology, including low SNRs, preferred orientation artifacts, and missing-wedge problems. The review is organized to follow the chronological workflow of cryoEM processing, beginning with early computational approaches that preceded the deep learning revolution. Subsequent sections examine automated particle picking using convolutional neural networks, computational solutions for preferred orientation bias through tools like spIsoNet and cryoPROS, and advanced denoising algorithms such as Topaz-Denoise. The survey then explores cryoET-specific applications including missing-wedge correction with IsoNet and subtomogram averaging through TomoNet's AI-driven approaches. The final technical sections cover the transformation of density maps into interpretable atomic models using sophisticated tools like ModelAngelo, DeepTracer, and CryoREAD, followed by considerations of visualization and graphics tools. The review concludes by examining future directions, particularly the potential integration of large language models and multi-agent systems that could further revolutionize structural biology workflows through enhanced automation and accessibility.

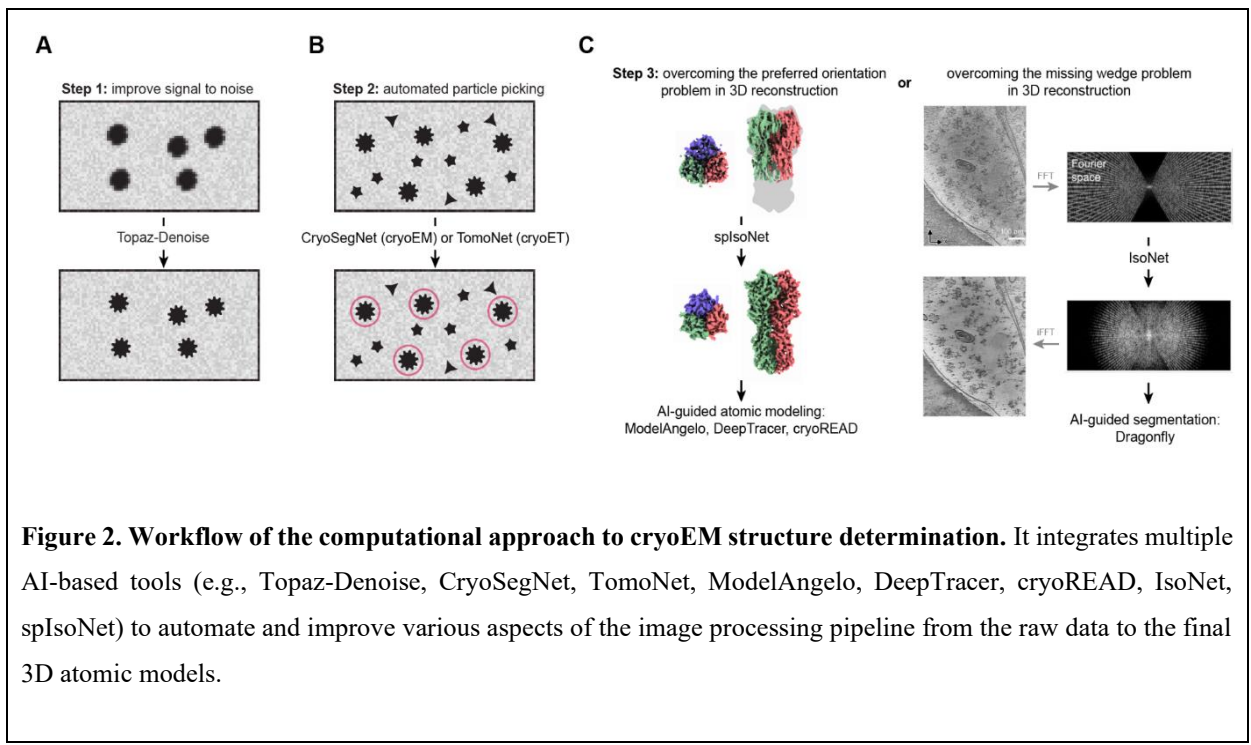

**Figure 2. Workflow of the computational approach to cryoEM structure determination.** It integrates multiple AI-based tools (e.g., Topaz-Denoise, CryoSegNet, TomoNet, ModelAngelo, DeepTracer, cryoREAD, IsoNet, spIsoNet) to automate and improve various aspects of the image processing pipeline from the raw data to the final 3D atomic models.

## 2. Early Image Processing Before the Advent of Deep Learning

Before examining specific AI applications, it is important to understand how these tools address the limitations of traditional cryoEM processing methods. Classical computational approaches relied primarily on statistical algorithms and iterative refinement procedures. For example, particle picking and orientation determination traditionally employed template matching and cross-correlation methods [9-15], requiring extensive manual validation and often failing with low-contrast or unusual particle orientations [16]. Three-dimensional reconstruction used expectation-maximization algorithms implemented in packages like FREALGN [15] and RELION [17], which, while successful, demanded significant computational resources and expert intervention.

The transition to deep learning methods represents a fundamental shift from rule-based algorithms to data-driven approaches. Rather than relying on predefined templates or statistical assumptions, deep learning models can automatically extract complex features from training data, enabling them to handle the inherent variability and noise challenges that limited traditional approaches. This has enabled solutions to previously intractable problems, such as the severe preferred-orientation problem and extremely low SNRs, while dramatically reducing cumbersome manual interventions.

## 3. Automated Particle Picking

As shown in Figure 2, particle picking is a critical step in the cryoEM workflow, where individual molecular complexes must be accurately identified and extracted from often noisy micrographs. Manual picking is time-consuming and subjective, thus motivating the development of automated particle pickers. Modern AI-based approaches use deep learning to distinguish true particles from contaminants and noise, while adapting to diverse protein shapes, sizes, and imaging conditions. The most significant models are Topaz and crYOLO, which use convolutional neural networks (CNNs), and CryoSegNet which uses a hybrid approach combining a U-net with a pre-trained foundation model.

Introduced by Bepler et al. [18], Topaz was among the first deep learning models to demonstrate the significant capabilities of automatic particle picking. The model uses a simple CNN-architecture, trained on sparsely labeled positive particles (real protein particles) and unlabeled data (no need for labeled negatives). CNNs are designed for image classification and are capable of object detection because of their exceptional ability to extract contextual features in images [19]. The Topaz pipeline can detect particles by preprocessing micrographs through Gaussian mixture normalization, which are then fed to a sliding-window CNN classifier that predicts particle locations. High-confidence detections are then refined through non-maximum suppression to extract precise coordinates. Unlike traditional methods, Topaz excels at picking challenging particles (small, asymmetric, or aggregated proteins) while maintaining low false-positive rates. Although representing a significant advancement in automatic particle picking, it still depends on the training data's label quality, it struggles with noisy micrographs, and it sometimes fails to generalize to unseen particle data. Despite these limitations, Topaz eliminates manual curation bias, retrieves more particles, and achieves high-resolution reconstructions (2.8–3.7 Å) with minimally labeled data [18].

crYOLO follows Topaz in automatic particle picking in cryoEM with a "You Only Look Once" (YOLO) deep learning model built on top of a CNN. Compared to a basic CNN — which is primarily an object classifier, not a detector — the benefit of using YOLO is that it incorporates additional object-detection layers into the CNN architecture. This enables YOLO to divide each micrograph into a grid and then predict particle centers, bounding boxes, and confidence scores within a single pass [20]. crYOLO thus improves upon Topaz by incorporating patch-processing modifications to handle small particles while employing data augmentation techniques, such as flipping and noise addition, to prevent overfitting. This speeds up particle detection, achieving a remarkable rate of approximately 5 micrographs per second on a single Graphics Processing Unit (GPU) while maintaining a high accuracy [21].

The benefits of the YOLO framework become even more apparent when examining crYOLO's underlying architecture and training approach. Built using 21-layer CNNs, 5 max-pooling layers, and trained on 45 diverse cryoEM datasets, the model also demonstrates zero-shot capabilities; i.e., accurately picking particles from completely unseen datasets. Retaining the CNN architecture enables it to learn contextual information and automatically avoid contamination and ice artifacts. This allows crYOLO to consistently outperform contemporary tools like RELION and EMAN2 by picking 29–30% more valid particles while reducing false positives. While crYOLO has notable limitations with low SNR datasets, the YOLO-based approach maintains high precision and recall even with moderately low SNR micrographs (AUC > 0.8). It demonstrates excellent centering accuracy (mean IOU > 0.8), ultimately leading to improved 3D reconstruction resolutions [21].

Despite crYOLO's impressive performance, researchers have continued to push the boundaries of automated particle picking by exploring novel architectural approaches that go beyond the YOLO framework. Recently, the CryoSegNet model introduced by Gyawali et al. [8] represents a significant advancement in automated protein particle picking by taking a hybrid approach. CryoSegNet integrates a specialized attention-gated U-net, trained on the large CryoPPP dataset, with the Segment Anything Model (SAM) to address the limitations of particle pickers such as crYOLO and Topaz. The U-net, enhanced with attention mechanisms, generates precise segmentation maps from noisy cryoEM micrographs, which are then processed by SAM's automatic mask generator to accurately localize protein particles. This hybrid AI approach outperforms other methods, achieving higher precision (0.792), F1-score (0.761), and Dice score (0.719) on a test dataset of 1,879 micrographs with 401,263 labeled particles [8]. CryoSegNet's ability to generalize across diverse protein types, coupled with its fine-tuning capability using predicted labels, enhances its robustness, making it a powerful tool for automating cryoEM data processing.

Beyond the improved segmentation metrics, CryoSegNet's AI-driven approach significantly enhances the resolution of 3D protein density maps, achieving an average resolution of 5.0 Å on the CryoPPP test dataset. This surpasses crYOLO (5.4 Å) and Topaz (5.2 Å). When tested on a larger set of micrographs from five protein types in EMPIAR, CryoSegNet achieved an average resolution of 3.3 Å, comparable to manually curated maps and up to 14% better than competing AI methods [8]. The model's efficiency stems from its symmetric encoder-decoder architecture, attention gates for prioritizing true particles, and post-processing to reduce false positives. Despite longer inference times due to these steps, CryoSegNet's superior accuracy and ability to handle diverse, noisy cryoEM data highlight its potential to streamline structural biology research, reducing reliance on labor-intensive manual or semi-automated particle picking methods [8].

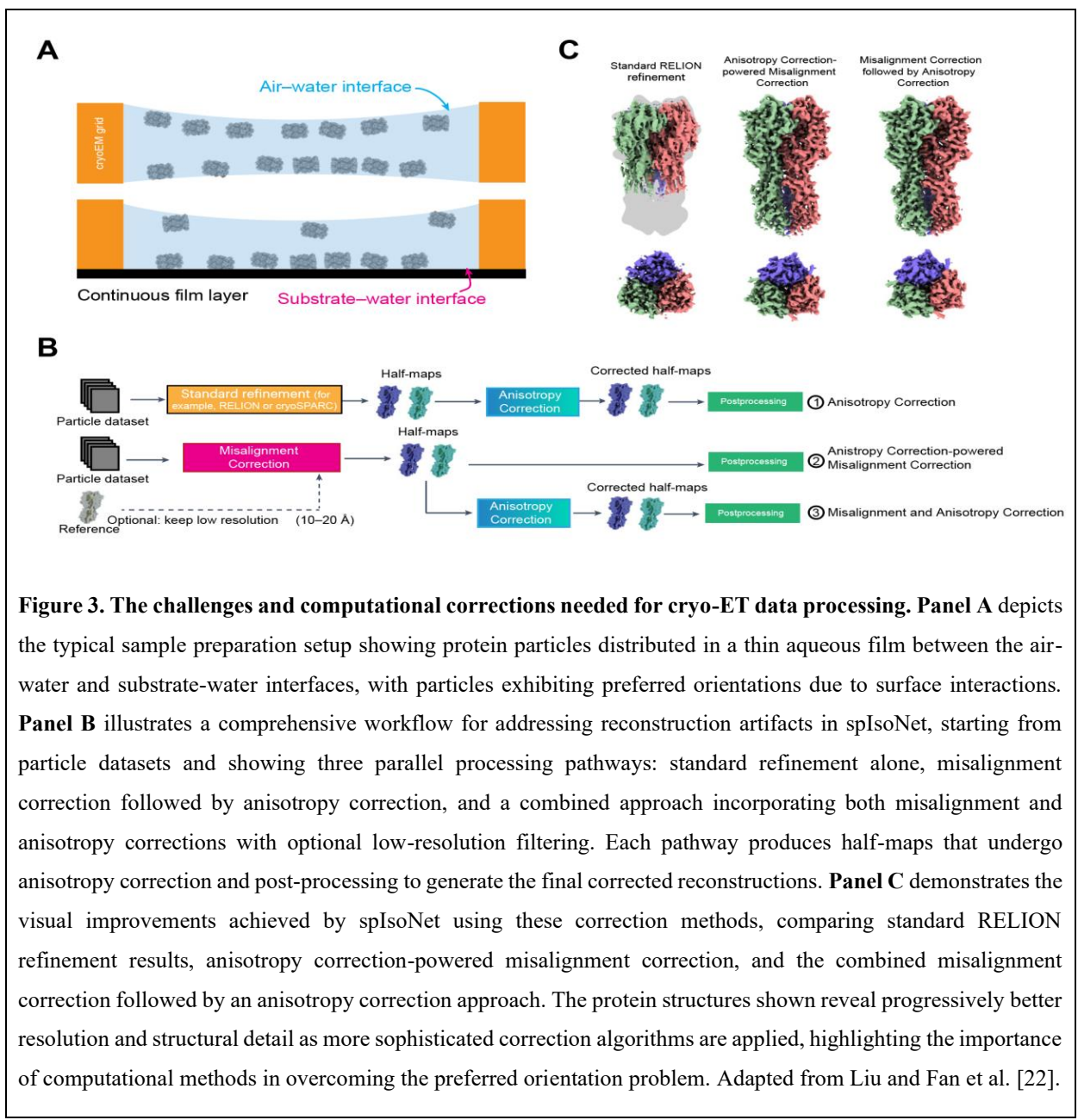

**Figure 3. The challenges and computational corrections needed for cryo-ET data processing. Panel A** depicts the typical sample preparation setup showing protein particles distributed in a thin aqueous film between the air-water and substrate-water interfaces, with particles exhibiting preferred orientations due to surface interactions. **Panel B** illustrates a comprehensive workflow for addressing reconstruction artifacts in spIsoNet, starting from particle datasets and showing three parallel processing pathways: standard refinement alone, misalignment correction followed by anisotropy correction, and a combined approach incorporating both misalignment and anisotropy corrections with optional low-resolution filtering. Each pathway produces half-maps that undergo anisotropy correction and post-processing to generate the final corrected reconstructions. **Panel C** demonstrates the visual improvements achieved by spIsoNet using these correction methods, comparing standard RELION refinement results, anisotropy correction-powered misalignment correction, and the combined misalignment correction followed by an anisotropy correction approach. The protein structures shown reveal progressively better resolution and structural detail as more sophisticated correction algorithms are applied, highlighting the importance of computational methods in overcoming the preferred orientation problem. Adapted from Liu and Fan et al. [22].

## 4. Single Particle CryoEM: Overcoming the Preferred-Orientation Problem

Even with the aforementioned advances in automated particle picking — from Topaz's template-based approach to crYOLO's YOLO framework and CryoSegNet's hybrid segmentation strategy — the preferred-orientation problem remains a major challenge for many biological specimens. This limitation affects all particle pickers as it stems not from computational deficiencies but from the physical behavior of particles during sample preparation. Ideally, particles in cryoEM samples should adopt a wide range of orientations within a thin layer of vitreous ice. However, as shown in Figure 3A, particles often adopt only a limited number of views due to uneven interactions with the Air-Water Interface (AWI) or support Film-Water Interface (FWI) during vitrification [23, 24]. This orientation bias leads to anisotropic three-dimensional reconstructions and artifacts such as skewed secondary structures, broken peptides, and distorted side chains (Figure 3B). These inaccuracies pose a significant obstacle to atomic modeling (Figure 3C) and undermine cryoEM's capacity for high-throughput structure determination and the informing of structure-guided drug design [25].

Traditional attempts to mitigate the preferred orientation problem have relied on biochemical or physical interventions during specimen preparation. Biochemical approaches include the use of surfactants to mask the AWI [26-28] or chemical modifications of the specimen [29] to reduce orientation bias. Physical methods, such as using support films [30-35], using time-resolved vitrification [36-40], collecting micrographs of thicker regions [41], or tilting the sample during acquisition [42, 43], can also diversify particle views. While these methods have been demonstrated to be effective, they are often specimen-specific, technically demanding, and require careful optimization that may inadvertently compromise specimen integrity.

In a fresh conceptual approach, Liu et al. [22] addressed solving the preferred orientation problem not by optimizing specimen preparation, but by improving the computational processing of anisotropic datasets using deep learning. They introduced single-particle IsoNet (spIsoNet), a self-supervised deep learning framework designed to correct angular anisotropy and particle misalignment without requiring any changes to specimen preparation. It consists of two modules, map anisotropy correction and anisotropy correction-powered misalignment correction, that together enhance reconstruction quality. The model uses abundant preferred-orientation views to infer information missing from under-represented orientations, effectively restoring isotropy in the angular distribution of particles.

spIsoNet was also shown to significantly improve alignment accuracy and map isotropy in reconstructions of several biological specimens, including ribosomes, β-galactosidases, and a hemagglutinin (HA) trimer dataset that had previously resisted high-resolution reconstruction using conventional methods. Specifically, it enabled a 3.5 Å reconstruction of non-tilted HA particles that was previously intractable without tilting, demonstrating its ability to overcome severe orientation bias. Additionally, the method generalizes to improving the reconstructions of preferentially oriented molecules in subtomogram averaging for cryoET, offering broader utility beyond single-particle analysis. Unlike earlier solutions, spIsoNet requires no specimen optimization, offering a robust, generalized computational solution to address the preferred orientation problem [44].

In a complementary advance, Zhang et al. [45] developed cryoPROS (PReferred Orientation dataset Solver), a computational approach designed to correct reconstruction misalignment caused by the preferred orientation problem by co-refining raw and auxiliary particles. CryoPROS also achieved improved isotropic reconstructions of a non-tilted HA-trimer dataset suffering from orientation bias, as well as other difficult experimental datasets. However, caution is prudent when choosing the initial reference model in generating auxiliary particles to avoid template bias introduced by relying on knowledge from other structures.

These computational strategies offer a convenient alternative to traditional specimen optimization workflows in addressing the preferred orientation problem. They may also be integrated with existing workflows and reduce the burden of extensive sample optimization. As deep learning methods continue to evolve, we anticipate and await the future deployment of more advanced deep learning architectures that will further enhance isotropic reconstruction, making cryoEM more robust, reproducible, and accessible for the broad structural biology community.

## 5. Denoising

Beyond orientation bias, another fundamental challenge limiting cryoEM efficiency is poor Signal-to-Noise ratios [18]. Like the preferred-orientation problem, a low SNR has traditionally required careful optimization of experimental conditions, yet computational approaches using deep learning have emerged as powerful alternatives that can address this limitation post-acquisition. SNRs in cryoEM can be as low as 0.11, making it difficult for automated particle pickers to identify protein particles. This particularly affects small proteins and non-globular structures with challenging orientations. Low SNRs can slow down the cryoEM pipeline due to bottlenecks in particle picking, classification, and 3D reconstruction. Furthermore, the push toward high-throughput facilities and even smaller protein targets has resulted in an acute need for powerful image enhancement methods.

Topaz-Denoise, introduced by Bepler et al. [46], builds upon the Topaz model [18] by integrating its deep neural networks trained with a denoising function. Denoise is a Noise2Noise framework based on thousands of micrographs from multiple detector systems. Trained on independent video frames — paired with noisy observations of the same signal, they circumvent the need for ground truth images. Topaz, which uses a U-net model, improves SNR by approximately 100-fold over the raw images and >2 dB over conventional filtering. It also works across different cameras and conditions, without dataset-specific retraining.

Topaz-Denoise is also capable of 75–90% shorter exposure times while maintaining image quality. This increases microscope throughput by 15–65% and yields thousands more exposures daily. For example, in a clustered protocadherin case study [46], Topaz-Denoise models achieved substantially improved performance compared to those trained on raw images, identifying 2.15 times more particles while generating a more complete sampling of difficult-to-detect orientations (such as top views). It revealed previously invisible particle orientations in clustered protocadherin, leading to discovery of a novel conformational state and 2.15x more particles for reconstruction. The software includes both 2D and 3D denoising capabilities and integrates with major cryoEM platforms, though the authors emphasize using denoised images only for visualization and particle identification, not direct reconstruction in order to avoid hallucination artifacts. Topaz-Denoise, along with spIsoNet and cryoPROS, are examples of successfully using AI to overcome physical limitations computationally rather than experimentally.

# 6. Cryogenic Electron Tomography

The computational solutions developed for single particle cryoEM have found natural extensions in cryoET, where similar challenges of noise and missing information are compounded by the complexities of three-dimensional cellular environments. Building on denoising approaches like Topaz-Denoise, cryoET tools have integrated AI to address both familiar problems and tomography-specific challenges such as the missing-wedge artifact.

The M software developed by Tegunov et al. [47] demonstrates how AI-enhanced approaches can achieve remarkable results in complex cellular contexts, such as resolving a chloramphenicol-bound 70S ribosome at 3.5 Å inside Mycoplasma pneumoniae cells, the first atomic-resolution structure of a drug-target complex in situ. A key AI-driven feature of M is its CNN-based denoising, which, like Topaz-Denoise, uses the Noise2Noise training regime to filter half-maps to local resolution during refinement iterations, thus mitigating overfitting artifacts in disordered regions like lipid bilayers. This AI approach, integrated with Warp and RELION, optimizes particle poses, corrects electron-optical aberrations, and models spatial deformations, enabling residue-level structural insights in complex cellular environments. By unifying frame and tilt-series processing, M's AI-enhanced pipeline significantly improves convergence speed and map quality, facilitating structural biology advancements.

IsoNet is another cryoET package that relies on deep learning to address the missing-wedge problem, which causes anisotropic resolution and hinders structural interpretation [6]. IsoNet uses a U-net-based neural network that has a symmetric encoder-decoder architecture, featuring 3D convolutional layers, strided convolutions, and skip-connections to reconstruct missing-wedge information and enhance SNRs. Trained on raw tomograms without subtomogram averaging, IsoNet iteratively refines subtomograms by mapping noisy, missing-wedge-affected inputs to corrected targets, incorporating simulated Gaussian noise models to mimic tomogram reconstruction artifacts. This approach enables IsoNet to produce tomograms with significantly reduced resolution anisotropy, thus improving structural interpretability across diverse biological samples.

IsoNet's deep learning framework is particularly innovative in its ability to simultaneously perform missing-wedge correction and denoising. The training process involves extracting subtomograms, applying a missing-wedge filter, and optionally adding noise to simulate real-world tomogram conditions, with noise levels gradually increased over iterations to ensure stable convergence. The U-net architecture, with three encoder and decoder blocks, uses 3x3x3 convolutional kernels and ReLU activations to extract complex features, while transpose convolutions in the decoder path restore full spatial dimensionality. Skip-connections preserve high-resolution details lost during downsampling, enhancing the network's ability to reconstruct fine structural features. By processing datasets including HIV particles, eukaryotic flagella, and neuronal synapses, IsoNet demonstrates superior performance, resolving intricate details such as lattice defects in HIV capsids and heptagon-containing clathrin cages, which are often obscured in traditional cryoET reconstructions [6].

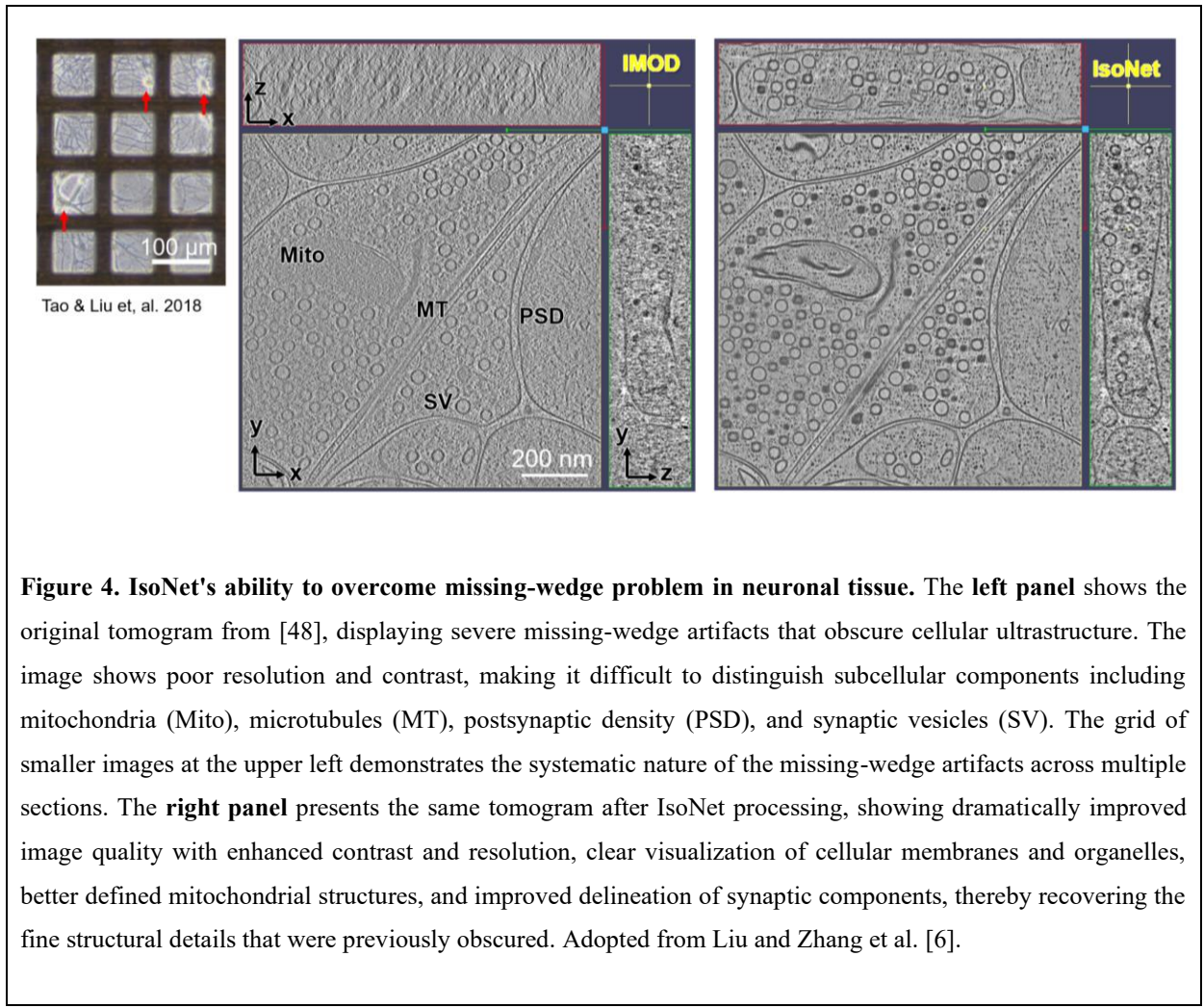

**Figure 4. IsoNet's ability to overcome missing-wedge problem in neuronal tissue.** The **left panel** shows the original tomogram from [48], displaying severe missing-wedge artifacts that obscure cellular ultrastructure. The image shows poor resolution and contrast, making it difficult to distinguish subcellular components including mitochondria (Mito), microtubules (MT), postsynaptic density (PSD), and synaptic vesicles (SV). The grid of smaller images at the upper left demonstrates the systematic nature of the missing-wedge artifacts across multiple sections. The **right panel** presents the same tomogram after IsoNet processing, showing dramatically improved image quality with enhanced contrast and resolution, clear visualization of cellular membranes and organelles, better defined mitochondrial structures, and improved delineation of synaptic components, thereby recovering the fine structural details that were previously obscured. Adopted from Liu and Zhang et al. [6].

The practical implementation of IsoNet, detailed in [6], showcases its robustness and accessibility, with the software and demonstration publicly available on GitHub. For instance, processing HIV tomograms from EMPIAR-10164 involved extracting 300 subtomograms, training two independent networks over 35 iterations, and averaging their predictions to achieve isotropic resolution, validated by 3D Fourier Shell Correlation (FSC). Similarly, ribosome and neuronal synapse datasets were processed with tailored parameters, demonstrating IsoNet's adaptability to varying sample complexities, as shown in Figure 4 for synaptic junctions among neuronal cells cultured on a cryoET grid. By overcoming the need for extensive manual intervention or subtomogram averaging, IsoNet's deep learning approach not only streamlines cryoET data analysis but also enhances the potential for high-resolution in situ structural biology, paving the way for broader applications in understanding molecular sociology within cellular contexts. More recently, IsoNet was further enhanced to overcome general orientation anisotropy problems in both cryoEM and cryoET by minimizing consistency and equivariance losses during AI training [22].

## 7. Subtomogram Averaging

The enhanced tomograms, as shown in Figure 4, produced by reconstruction tools like IsoNet create opportunities for the more sophisticated analysis of individual macromolecular complexes through SubTomogram Averaging (STA). Recognizing this potential, Wang et al. [7] introduced TomoNet, a comprehensive software pipeline that applies deep learning specifically to the challenges of particle identification and averaging within tomographic volumes. TomoNet's "AI AutoPicking" module leverages a U-net convolutional neural network to perform voxel-wise binary classification, identifying particle centers within tomograms. This supervised machine learning approach requires only 1–3 tomograms with known particle coordinates as ground truth, from which subtomograms and segmentation maps are generated for training. The trained model efficiently predicts particle coordinates across entire datasets, reducing manual intervention and enabling rapid processing on GPUs, typically within minutes per tomogram. By integrating with tools like RELION for high-resolution refinement, TomoNet's deep learning capabilities enhance the accuracy and efficiency of STA, achieving near-atomic resolutions, as demonstrated with the HIV Gag hexamer at 3.2 Å [7].

TomoNet's deep learning-based particle picking complements its geometric "Auto Expansion" module, offering versatility for both lattice-organized and scattered particles, such as ribosomes. Unlike traditional template matching, AI AutoPicking does not rely on prior lattice knowledge, learning 3D features directly from training data, which makes it adaptable to diverse sample types. The module's efficiency is evident in its application to datasets including HIV virus-like particles, eukaryotic axonemes, and bacterial S-layers, where it successfully handled low-contrast and imperfect lattices. While AI AutoPicking may pick fewer particles than Auto Expansion on highly flexible lattices, its speed and reduced need for manual input — saving approximately 5–15 minutes per tomogram — make it a powerful tool for large-scale cryoET projects. TomoNet's modular design ensures seamless integration of its deep learning advancements with existing cryoET workflows, thereby lowering barriers for researchers and facilitating high-resolution in situ structural studies [7].

## 8. Density Maps to Atomic Models: Automated Protein Model Building

Having established how deep learning transforms the acquisition and processing stages of both single-particle cryoEM and cryoET, our focus now turns to the final stages that interpret the resulting density maps to produce atomic models. While earlier sections focused on improving data quality and particle identification, the modeling stage is where enhanced density maps are transformed into interpretable biological structures. AI has demonstrated significant capabilities for this final but crucial step, with tools that can automatically build protein and nucleic acid models directly from cryoEM density maps.

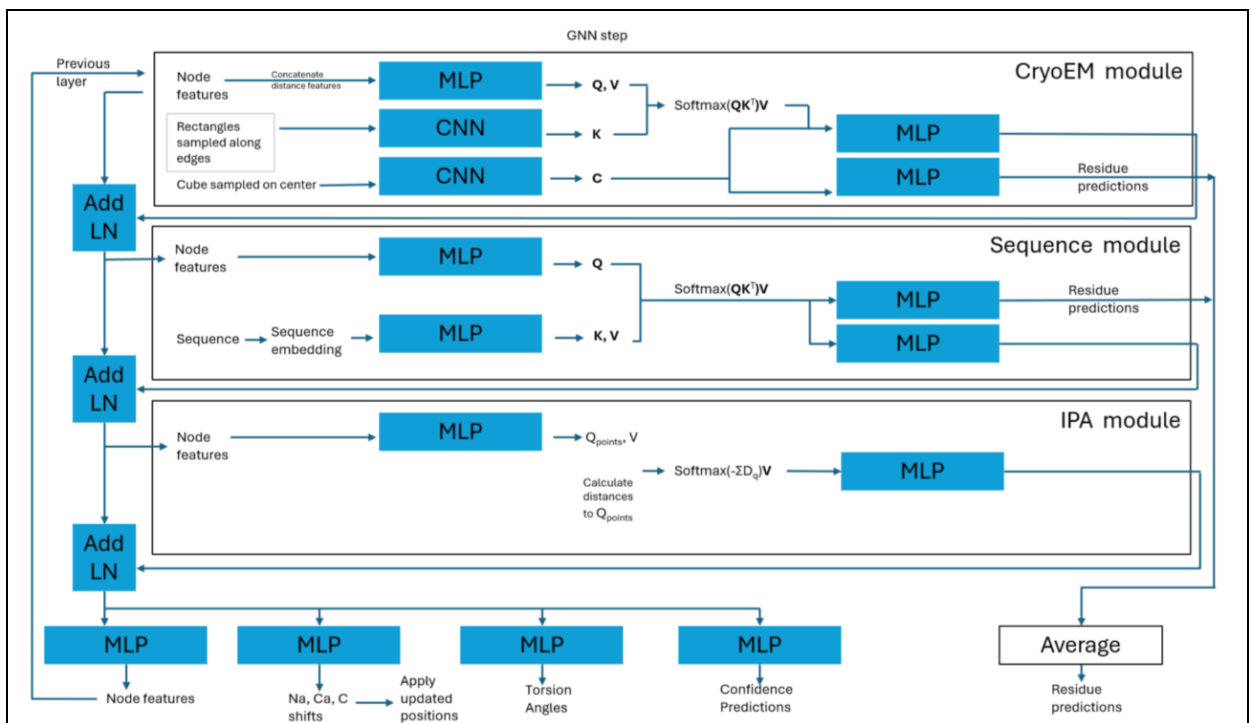

**Figure 5. The ModelAngelo pipeline [49] for automated atomic model construction from cryoEM density maps**. The computational architecture features multiple specialized modules that contain Multi-Layer Perceptrons (MLPs) and CNNs. The modules comprise a cryoEM module that processes density features through CNNs, a Sequence module that incorporates protein sequence information via embeddings and evolutionary data, and an Interface Prediction Algorithm (IPA) module that calculates the spatial relationships between predicted residues. The pipeline integrates structural features extracted from the density map with sequence-based information through iterative refinement steps, ultimately generating comprehensive atomic coordinates, residue assignments, and confidence predictions for the final protein structure model.

Three prominent examples demonstrate the range and sophistication of these AI-driven modeling approaches. ModelAngelo [49] is a machine-learning tool for automated atomic model building in cryoEM maps. It combines cryoEM density with protein sequence and structural information through a Graph Neural Network (GNN) illustrated in Figure 5. It is capable of both building protein models and identifying unknown protein sequences, using predicted amino acid probabilities from hidden Markov model searches. For nucleic acids, it constructs backbone models with human-level accuracy. Another model, DeepTracer [50], focuses on protein structure determination from cryoEM maps. It uses deep learning to trace polypeptide chains and assign sequences, though it has limited support for nucleic acids. CryoREAD [51] is a model that specializes in de novo DNA/RNA structure modeling from cryoEM maps (2.0–10.0 Å resolution). It employs deep learning to detect phosphate, sugar, and base positions, and then assembles them into full atomic models. It achieves >85% backbone accuracy and outperforms previous methods for nucleic acids, including complex structures such as SARS-CoV-2 RNA-protein complexes. While ModelAngelo and DeepTracer prioritize protein modeling with varying degrees of nucleic acid support, CryoREAD fills a critical gap by automating nucleic acid structure determination, which is otherwise manually intensive. All three tools enhance objectivity and efficiency in cryoEM structure modeling.

## 9. Graphics

While AI has automated much of the technical processing pipeline, the visualization and interpretation of structural data remains a critical human-centered aspect of the cryoEM and cryoET workflows. Graphics tools, particularly segmentation software and molecular visualization platforms, serve as the essential interface between the AI-generated results and our biological understanding. These tools must handle data across the entire pipeline, from raw tomograms requiring segmentation to final atomic models needing validation and presentation. Graphics in cryoEM are essential for transforming raw density maps into interpretable biological structures. Segmentation tools (cryoEM/ET software such as IMOD [52] or EMAN2 [53] enable 3D volume segmentation to isolate and label macromolecular complexes, membranes, or cellular features from tomograms or density maps. They support threshold-based masking, machine-learning-aided boundary detection, and manual tracing, which identifies biologically relevant regions for further analysis. UCSF Chimera [54] and ChimeraX [55] provide advanced molecular visualization with capabilities for volume rendering, surface representation, and atomic model fitting into cryoEM maps. Features such as lighting adjustments, transparency controls, and coloring by properties (e.g., B-factors) enhance interpretability, while plugins integrate tools for flexible fitting, density segmentation, and animation. Together, these tools bridge raw data and structural insights, offering both precision and flexibility for macromolecular complexes at near-atomic resolution.

## 10. The Future of AI in CryoEM

The transformative advances in cryoEM and cryoET, while revolutionary, represent only the beginning of what is possible as AI research continues to evolve. The tools surveyed, from particle picking to modeling, have largely focused on specific technical challenges within established workflows. The current wave of State-Of-The-Art (SOTA) AI research revolves around Large Language Models (LLMs), and the future evolution of AI in cryoEM may be driven by LLMs. Modern LLMs are decoder-transformers trained on huge text corpora. The text data are reduced to tokens, with a select few receiving focus (through attention mechanisms) that enables the LLM to understand and generate human-like language [56]. LLMs and Vision-Language Models (VLMs) — which replace text tokens with image tokens — parallel AI advances in cryoEM. For example, just as tools like CryoSegNet or ModelAngelo predict protein structures or particle positions, existing LLMs/VLMs are designed to predict text or detect objects in an image. Moreover, many SOTA models are trained on existing models through a process called fine-tuning, and new tools in cryoEM could fine-tune an LLM with cryoEM data. The adaption of foundation models to cryoEM has already been successfully demonstrated with CryoSegNet, whose hybrid approach combines a novel U-net with an existing foundation model. Furthermore, the attention mechanism of LLMs/VLMs would be especially useful in particle picking, where context-aware reasoning could improve the detection of low-SNR particles.

The success of AlphaFold [57] underscores the transformative potential of attention-based architectures in structural biology. AlphaFold's Evoformer module employs a attention mechanism to infer spatial relationships between amino acids, enabling accurate protein structure prediction even without direct experimental data. Similarly, CryoSegNet uses attention gates in its U-net to prioritize relevant features in noisy cryoEM micrographs, enhancing particle segmentation [8]. However, modern LLMs like GPT-4 or Vision Transformers (ViTs) leverage attention mechanisms such as multi-head self-attention, hierarchical attention, and graph head attention [58], capabilities that could surpass CryoSegNet's localized attention by integrating global context (e.g., correlating particle orientations across tomograms). AlphaFold's reliance on evolutionary data also highlights another frontier: synthetic data generation. CryoEM models already confront limited data problems, but augmenting datasets with simulated particle distributions or micrograph noise profiles may help improve generalization, not dissimilar to AlphaFold's use of multiple sequence alignments to fill knowledge gaps.

Beyond individual AI tools, there is significant potential in reimagining independent LLM tools into multi-agent-LLM systems [59]; i.e., networks of autonomous, collaborating agents fine-tuned to each task in the cryoEM workflow. Such a system would deploy acquisition agents, preprocessing agents, and validation agents that dynamically adapt workflows based on real-time feedback. Acquisition agents could work across the pipeline, adjusting microscope parameters such as defocus and tilt angles upon detecting orientation bias using spIsoNet's predictions, while denoising agents would selectively apply Topaz-Denoise only to low-SNR micrographs flagged by quality-control modules. Reconstruction swarms could parallelize subtomogram averaging across GPUs, with validator agents cross-checking outputs against ModelAngelo's atomic models to prevent overfitting. Such agent-based systems would address two critical gaps in current workflows: end-to-end automation and adaptive problem-solving. For the former, agents could bridge disconnected tools by passing CryoSegNet's particle picks directly to IsoNet for missing-wedge correction. Furthermore, agents are adaptive; reinforcement learning agents could co-refine auxiliary particles using cryoPROS and correct misalignments with spIsoNet in iterative loops, thus overcoming the limitations of static processing pipelines.

## 11. Conclusions

In this review we examined how the integration of deep learning technologies has fundamentally transformed cryoEM and cryoET workflows by addressing critical bottlenecks that previously required extensive manual intervention and specimen optimization. The transition from rule-based algorithms to data-driven approaches has enabled solutions to previously intractable problems, including severe preferred orientation bias, extremely low SNRs, and missing-wedge artifacts in tomography. Key advances span the entire pipeline: improved automated particle picking, computational correction of orientation bias, novel denoising techniques, and automated atomic model building that achieves near-atomic resolution reconstructions with minimal manual intervention. These AI-enhanced approaches have demonstrated remarkable achievements across diverse biological systems — from HIV virus-like particles to in situ ribosomal complexes — significantly expanding the scope and efficiency of structural biology research. The success of these computational solutions represents a paradigm shift from optimizing experimental conditions to post-acquisition computational enhancement, making high-resolution structural analysis more accessible, reproducible, and efficient.

      Looking forward, the field of cryoEM can greatly benefit through the integration of LLM and agent-based systems. In and of itself, incorporating LLMs could streamline cryoEM workflows by automating particle picking, guiding image processing, and enhancing existing models with synthetic data. By integrating the models throughout the cryoEM pipeline into a multi-agent LLM, one can expect that enabling agent cooperation may improve the system as a whole. Multimodal LLMs, such as GPT-4o, already show improved performance by combining data across different modalities [59, 60]. Overall, LLMs and agents already exemplify how AI can extract meaningful patterns from noisy, high-dimensional data — a challenge central to cryoEM reconstruction. Their success elsewhere hints at opportunities for similar models trained on cryoEM datasets to automate tasks throughout the cryoEM pipeline.